%% file: main.tex
\documentclass[acmsmall,nonacm]{acmart}
\acmJournal{TOSEM}

\setcopyright{none}
\acmDOI{}

\usepackage{xcolor}
\usepackage{booktabs}
\usepackage{tabularx}
\usepackage{multirow}
\usepackage{tcolorbox}
\usepackage{graphicx}
\usepackage{amsmath}
\usepackage{siunitx}
\usepackage{url}
\usepackage{xspace}
\hypersetup{
  hidelinks,
  pdftitle={SWE-Manager: Selecting and Synthesizing Golden Proposals Before Coding},
  pdfauthor={Boyin Tan, Haoning Deng, Junyuan Zhang, Junjielong Xu, Pinjia He, Youcheng Sun}
}

\AtBeginDocument{%
  }

\newcommand{\swemanager}{{SWE-Manager}\xspace}
\newcommand{\modelName}{SWE-Manager-8B\xspace}
\newcommand{\frameworkName}{P2A\xspace}
\newcommand{\proposalagent}{\textsc{Proposal Agent}\xspace}
\newcommand{\technicalmanager}{\textsc{Technical Manager}\xspace}
\newcommand{\implementationagent}{\textsc{Implementation Agent}\xspace}

\begin{document}

\title{\swemanager: Selecting and Synthesizing Golden Proposals Before Coding}

\author{Boyin Tan}
\authornote{Boyin Tan and Haoning Deng contributed equally.}
\email{Boyin.Tan@mbzuai.ac.ae}
\affiliation{%
  \institution{Mohamed bin Zayed University of Artificial Intelligence}
  \city{Abu Dhabi}
  \country{United Arab Emirates}
}

\author{Haoning Deng}
\authornotemark[1]
\email{haoning.deng@aalto.fi}
\affiliation{%
  \institution{Aalto University}
  \country{Finland}
}

\author{Junyuan Zhang}
\email{junyuan.zhang@connect.hku.hk}
\affiliation{%
  \institution{The University of Hong Kong}
  \country{China}
}

\author{Junjielong Xu}
\email{junjielongxu@cuhk.edu.cn}
\affiliation{%
  \institution{The Chinese University of Hong Kong, Shenzhen}
  \country{China}
}

\author{Pinjia He}
\email{hepinjia@cuhk.edu.cn}
\affiliation{%
  \institution{The Chinese University of Hong Kong, Shenzhen}
  \country{China}
}

\author{Youcheng Sun}
\authornote{Youcheng Sun is the corresponding author.}
\email{Youcheng.Sun@mbzuai.ac.ae}
\affiliation{%
  \institution{Mohamed bin Zayed University of Artificial Intelligence}
  \city{Abu Dhabi}
  \country{United Arab Emirates}
}

\renewcommand{\shortauthors}{Tan et al.}

\newcommand{\modify}[1]{\textcolor{red}{#1}}
\newcommand{\revised}[1]{{\color{blue}#1}}
\newcommand{\eg}{e.g.,}
\newcommand{\commentout}[1]{}

\tcbset{
    colback=gray!10,    %
    colframe=black,     %
    boxrule=0.5mm,      %
    arc=3mm,            %
    auto outer arc,
    width=0.95\linewidth,   %
    left=5pt,          %
    right=5pt,          %
    boxsep=5pt,         %
    before=\vskip10pt,  %
    after=\vskip10pt    %
}

\input{sec/000_Abstraction}


\maketitle

\input{sec/001_Introduction}
\input{sec/003_Pilot_Study}
\input{sec/004_Method}
\input{sec/005_Experiment}
\input{sec/006_Conclusion}

\bibliographystyle{ACM-Reference-Format}
\bibliography{ref}

\end{document}

%% file: sec/000_Abstraction.tex
\begin{abstract}

    Large language model (LLM) research in software engineering has largely focused on tasks such as code generation and bug repair. An earlier step in real development workflows, proposal selection, remains underexplored. Teams often draft multiple candidate proposals for the same issue and decide one to guide implementation. This proposal-selection problem is challenging because multiple candidates may be technically plausible, and the best choice depends on issue context, regression risk, fix depth, maintainability, and product or workflow constraints. Yet it remains unclear whether proposal selection contains reusable reasoning patterns that LLMs can learn, rather than only instance-specific maintainer preferences.

    We address this question by first conducting a manual study of public GitHub issue discussions. Our study shows that maintainer proposal selection is context-conditioned, guided by recurring high-level criteria, and often useful for refining the final implementation plan. Based on these findings, we introduce \swemanager, a joint selection-and-synthesis approach that compares candidate proposals under the issue context, selects the best proposal, explains the choice, and synthesizes a golden proposal for implementation. We instantiate \swemanager as \modelName, an 8B model trained from Qwen3-8B using supervised fine-tuning and reinforcement learning.

    We evaluate \modelName on two capabilities: proposal selection and downstream usefulness of the synthesized golden proposal. On the SWE-Lancer Manager benchmark, \modelName achieves 53.21\% selection accuracy and a 57.75\% earned rate (\$152{,}750), outperforming GPT-5, which achieves 44.15\% accuracy and a 37.75\% earned rate (\$99{,}875). To evaluate whether the golden proposal can guide implementation, we introduce \frameworkName, a proposal-first, implementation-later workflow, and evaluate it on the SWE-Lancer IC benchmark. \frameworkName with \swemanager achieves a 55.6\% issue-resolution rate, matching the performance of \frameworkName with GPT-5 and improving over the mini-SWE-agent by 7.1\%. These results suggest that proposal selection and synthesis are learnable and practically useful capabilities for LLM-based software engineering systems.

\end{abstract}

%% file: sec/001_Introduction.tex
\section{INTRODUCTION}

Selecting an appropriate proposal is a recurring challenge throughout the software engineering life cycle~\cite{7961467, Magabaleh2024SystematicRO, Cunha2016DecisionMakingIS, Kumar2022MCDMBF}.
Early in a project, teams decide among architectural styles, technology stacks, and design alternatives~\cite{Vliet2016DecisionMI}. During development and maintenance, they evaluate competing proposals for bug fixes, refactorings, and feature implementations~\cite{11029753}.
Such decisions are rarely determined by technical feasibility alone; they often require balancing correctness, regression risk, maintainability, compatibility, delivery cost, and organizational constraints~\cite{Vliet2016DecisionMI,7961467,Cunha2016DecisionMakingIS,Kumar2022MCDMBF}.

Recent advances in large language models (LLMs) have substantially improved many software engineering tasks, including code generation~\cite{Guo2024DeepSeekCoderWT, Rozire2023CodeLO, lozhkov2024starcoder} and automated debugging~\cite{10.1145/3715754, Lin2025SEAgentST, 10.1109/ICSE55347.2025.00169}.
To further improve issue resolution, many systems generate multiple candidate patches and select one after implementation~\cite{10.1145/3715754, Antoniades2024SWESearchES}.
Such patch-level selection is costly: every candidate must be implemented in full and validated in a fresh copy of the environment, because agent actions modify the environment in place, and its states cannot be reliably saved, copied, or rolled back~\cite{Zainullina2025GuidedSearch}.
Selecting among candidate proposals before implementation avoids this cost, since the candidates are compared as natural-language plans and only the chosen one is implemented.
\input{input/Task_intro}
Despite this advantage, proposal selection before implementation has received far less attention than patch-level selection.
SWE-Lancer~\cite{Miserendino2025SWELancerCF} formalizes this setting as the \textit{Manager task}, where a manager reviews competing proposals for the same issue and selects the best one.
As illustrated in Fig.~\ref{fig:task_intro}, proposals for a real GitHub issue contain three parts: (1) a problem restatement, (2) a root-cause analysis, and (3) a proposed solution. In practice, all proposals may suggest plausible implementation directions.
A manager therefore needs to compare the features of proposals under the issue constraints rather than simply determine whether each proposal can address the issue.

Although SWE-Lancer~\cite{Miserendino2025SWELancerCF} formulated this problem and provides verified data for benchmarking proposal selection ability, it does not explore the process and rationale behind selection.
As a result, it remains unclear whether maintainer proposal selection contains reusable reasoning patterns that a model can learn, rather than only instance-specific preferences.
We therefore conduct a manual study of public GitHub issue discussions to characterize proposal selection in practice.

Our study reveals three properties of maintainer proposal selection. \textbf{First, maintainer choices are guided by recurring high-level rationales.}
Although each issue has its own context, maintainer selections are not arbitrary preferences.
Across cases, the reasons for preferring one proposal over others repeatedly appeal to engineering considerations such as risk and safety, fix depth, maintainability, and product or workflow expectations.
This regularity suggests that proposal selection contains learnable reasoning patterns: a selected-proposal label records only the final choice, while the accompanying rationale reveals the criteria behind that choice.
\textbf{Second, proposal selection is a context-conditioned comparison among candidate proposals.}
Maintainers compare candidate proposals under the issue context, where the relevant criteria and their relative importance emerge from the issue constraints and from the differences among proposals.
This makes the task different from a rubric-judge evaluation, in which a maintainer would apply a fixed scoring guide to each proposal in isolation and then choose the highest-scoring one.
For example, a minimal fix may be preferred when regression risk dominates, whereas a deeper root-cause fix may be favored when a narrow patch would leave the system fragile.
This observation motivates modeling proposal selection as a joint comparison over the issue and all candidate proposals, rather than as independent rubric-based scoring.
\textbf{Third, proposal discussion helps formulate the final implementation proposal.}
Proposal authors and maintainers often discuss and compare competing proposals.
These discussions can provide maintainers with selection cues and can refine the selected direction by clarifying assumptions, narrowing scope, rejecting risky steps, or incorporating useful observations from competing proposals.
This observation motivates treating proposal selection as a synthesis task: the model should not only return a selected proposal index, but also produce a \emph{golden proposal} that turns the selected direction and useful cross-proposal observations into implementation-facing guidance.

Building on those findings, we design \swemanager. We hypothesize that proposal selection is learnable from large-scale maintainer-selection data, and we operationalize this hypothesis with two design requirements: \swemanager compares the issue and all candidate proposals jointly, and it synthesizes a \emph{golden proposal} that turns the selected direction and useful cross-proposal observations into a refined implementation plan.
The model also explains its choice, making the learned decision inspectable rather than returning only a proposal index.
To verify this hypothesis, we instantiate \swemanager as \modelName, an 8B model trained from Qwen3-8B~\cite{Yang2025Qwen3TR} using supervised fine-tuning and reinforcement learning.

Evaluating \modelName requires testing two capabilities: whether it can select the best proposal among candidates, and whether its synthesized golden proposal can guide effective implementation.
The SWE-Lancer Manager benchmark~\cite{Miserendino2025SWELancerCF} directly evaluates the first capability, so we use it to measure \modelName's proposal-selection performance.
However, this benchmark does not evaluate the downstream usefulness of the synthesized golden proposal.
To evaluate this second capability, we use the synthesized golden proposal to guide an agent in generating a patch and use the patch's issue-resolution result as a proxy metric for proposal effectiveness.
Because \modelName does not itself generate candidate proposals or code patches, we introduce \frameworkName, an evaluation workflow inspired by real development workflows in which candidate proposals are drafted, reviewed, and selected for implementation.
In \frameworkName, proposal agents first generate multiple candidate proposals, \swemanager selects and synthesizes a golden proposal, and an implementation agent follows the golden proposal to produce a patch.
\frameworkName is not intended as another multi-agent framework: it keeps the proposal agents and the implementation agent fixed, so that differences in issue resolution can be attributed to the Technical Manager.
We then evaluate this workflow on the SWE-Lancer IC benchmark~\cite{Miserendino2025SWELancerCF} to test whether the golden proposal can help the implementation agent resolve the issue.
The results support both capabilities. On the SWE-Lancer Manager benchmark, \modelName achieves 53.21\% selection accuracy and a 57.75\% earned rate (\$152{,}750), outperforming GPT-5~\cite{openai_gpt5}, which achieves 44.15\% selection accuracy and a 37.75\% earned rate (\$99{,}875). On the SWE-Lancer IC benchmark, \frameworkName with \swemanager achieves a 55.6\% pass rate, matching the performance of GPT-5 and improving over the mini-SWE-agent~\cite{yang2024sweagent} baseline without proposal guidance by 7.1\%, suggesting that the synthesized golden proposal can provide useful guidance for downstream issue resolution.

In summary, our contributions are as follows:
\begin{itemize}
  \item We conduct a manual study of real-world maintainer proposal-selection behavior in GitHub issue discussions. The study formulates three insights: (1) maintainer choices are guided by recurring high-level rationales, (2) proposal selection is context-conditioned and comparative, and (3) proposal discussion helps formulate the final implementation proposal.
  \item We introduce \swemanager, an approach for selecting and synthesizing proposals before coding, and present \modelName, an 8B model instantiated from this approach. On the SWE-Lancer Manager benchmark, \modelName achieves 53.21\% accuracy and a 57.75\% earned rate (\$152{,}750 earned), surpassing GPT-5 by 9.06\% in accuracy and 20.00\% in earned rate, yielding \$52{,}875 more reward.
  \item We design \frameworkName, a proposal-first, implementation-later evaluation workflow, to evaluate whether the synthesized golden proposal is useful in end-to-end issue resolution on the SWE-Lancer IC benchmark. \frameworkName with \swemanager achieves a 55.6\% pass rate, matching the performance of P2A with GPT-5, and improves over the baseline by 7.1\%.
\end{itemize}

%% file: input/Task_intro.tex
\begin{figure*}[t]
  \centering
  \includegraphics[width=1\textwidth]{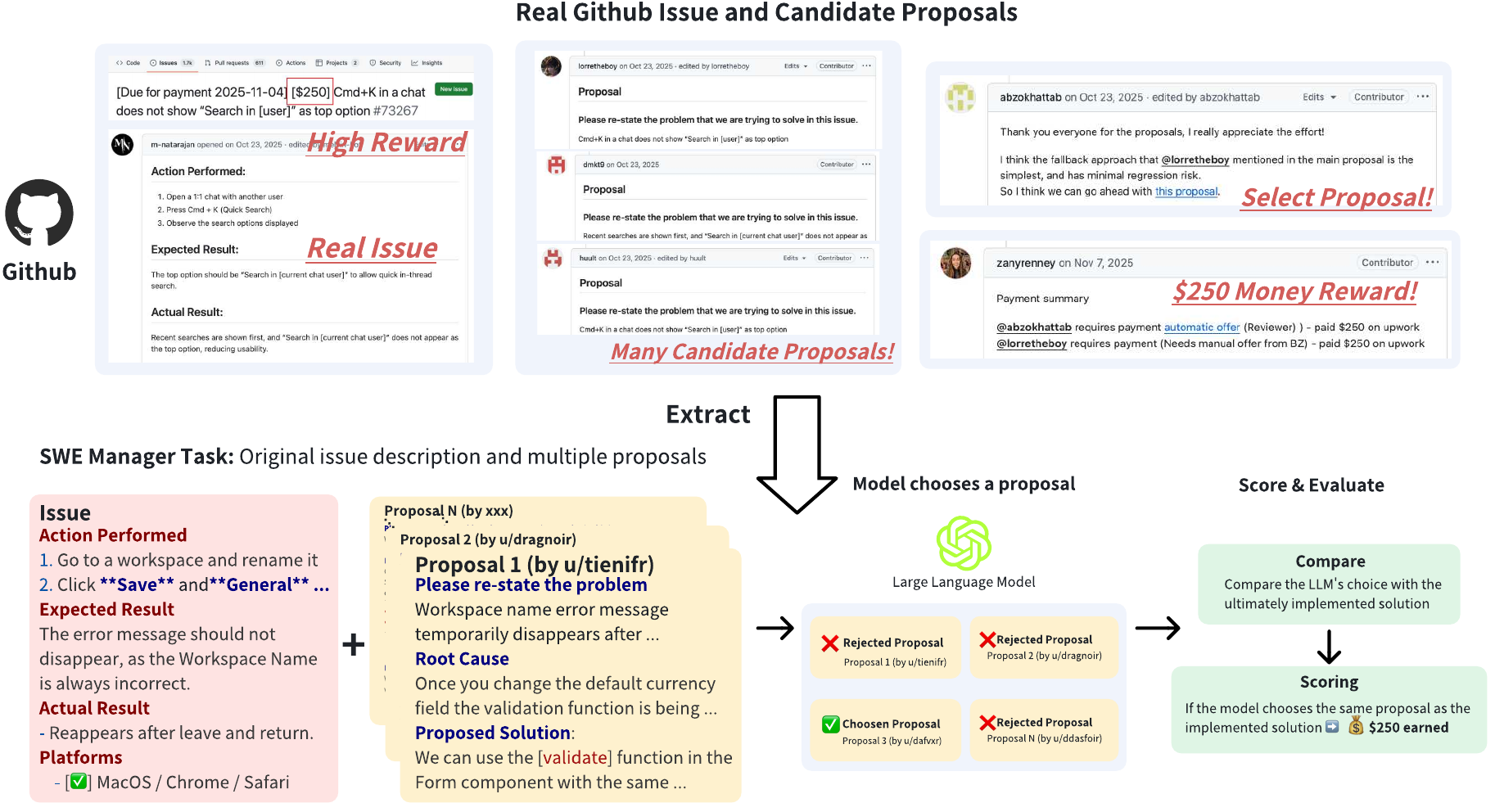}
  \Description{An example issue and competing implementation proposals form the input to a technical manager. The manager selects one proposal, and the selection is compared with the recorded reference choice.}
  \caption{Manager task overview. A real GitHub issue is paired with multiple independently written candidate proposals. The model plays the role of a technical manager: it reviews the candidates, rejects less suitable options, and selects the proposal that would be implemented. Performance is evaluated by comparing the selected proposal to the proposal that best matches the approach ultimately merged into the repository; in reward-based settings, a correct match earns the associated payout for that issue.}

  \label{fig:task_intro}
\end{figure*}

%% file: sec/003_Pilot_Study.tex
\section{MANUAL STUDY}
\subsection{Data Collection}
\label{subsec:data_collection}
To uncover the underlying principles of proposal selection, we follow the data collection protocol used by SWE-Lancer~\cite{Miserendino2025SWELancerCF} and construct a proposal-selection corpus from bounty-backed issues in open-source repositories hosted on GitHub~\cite{GitHub2026}.
For each issue, we first identify the user who ultimately received the bounty and then trace that user back to the corresponding proposal in the issue thread. We retain an issue only if it satisfies four criteria: (1) the issue is publicly accessible and has an associated bounty, (2) at least two candidate proposals are available, (3) each retained proposal follows the complete proposal structure with \texttt{Problem Restate}, \texttt{Root Cause Analysis}, and \texttt{Solution Suggest}, and (4) the reviewer provides clear selection reasons for the chosen proposal. We discard ambiguous threads where the paid user cannot be matched to a proposal, where fewer than two complete proposals are available, where the reviewer does not provide a selection reason, or where the selected proposal is only inferable post hoc from the final patch.

This process yields 2{,}852 issues and 7{,}500 candidate proposals, with an average of 2.63 proposals per issue. The resulting corpus provides naturally occurring examples of comparative proposal selection: each instance pairs one issue with at least two plausible proposals and a bounty-anchored selection outcome.
We use this corpus as the population for our manual study.

\subsection{Study Design}
Given the 2{,}852-issue corpus, we determine the required sample size using standard statistical techniques~\cite{Krejcie1970DeterminingSS}, following prior empirical studies~\cite{He2018CharacterizingTN,Gao2017ToTO}. Specifically, the sample size is computed as a function of the desired margin of error, confidence level, and population size. Setting the margin of error to $\pm 5\%$ and the confidence level to $95\%$, we obtain a required sample size of 339.

We designed the manual study as an exploratory coding study to understand how maintainers review competing proposals and decide which one should guide the repair.
We first reconstructed the decision episode in each sampled issue thread.
For each issue, the annotators recorded the issue context, the complete candidate proposals available before selection, the selected proposal, the maintainer comments that motivated the choice, and the discussion among maintainers, reviewers, and proposal authors.
This protocol allowed us to observe both the selection decision itself and the discussion process around that decision.

During the first coding pass, the annotators wrote free-text notes about what happened in each decision episode.
These notes captured three aspects of the process: what high-level rationale supported the final selection, how maintainers and proposal authors compared competing proposals, and how discussion before or after the decision shaped the final implementation direction.
Based on these notes, we coded each issue along three dimensions.
First, we assigned one primary selection rationale to capture the high-level engineering concern that best explained the choice.
After reviewing these rationale notes across cases, the authors found that the selection reasons repeatedly referred to a small set of high-level engineering concerns, and we merged semantically similar reasons into a compact rationale codebook.
Second, a case was labeled as a comparative decision when the maintainer's choice was supported by contrasts among candidate proposals' features under the issue context, rather than by judging a proposal in isolation.
Third, a case was labeled as containing refinement evidence when the thread showed proposal discussion shaping the final implementation direction, including clarification of assumptions, narrowing of scope, comparison among proposal authors, or incorporation of useful observations from competing proposals.

After the codebook was constructed, the first and second authors independently applied it to all sampled issues.
For each issue, they finalized the comparative-decision label, the primary selection rationale, and the refinement-evidence label.
Because the primary rationale is the most subjective multi-class label, we report its pre-reconciliation agreement: the two annotators agreed on the primary selection rationale for 312 of the 339 sampled issues (92.0\%).
They then reconciled all disagreements and finalized the labels used in the analysis.

\subsection{Manual Study Findings}
\label{subsec:reasons}

\textbf{\textit{Finding 1: Maintainer choices are guided by recurring high-level rationales.}}
The primary rationales cluster into three dominant families and one residual family (Table~\ref{tab:proposal_selection_rationales}): \emph{Risk and Safety} (145/339, 42.8\%), \emph{Fix Depth} (99/339, 29.2\%), \emph{Maintainability} (55/339, 16.2\%), and \emph{Others} (40/339, 11.8\%).
\emph{Risk and Safety} is the most frequent family: maintainers often preferred proposals that made smaller and more verifiable changes, constrained the fix scope, reduced regression risk, handled edge cases more robustly, or offered a time-critical mitigation.
\emph{Fix Depth} captures cases where maintainers favored proposals that addressed the underlying root cause or remediated a broader failure pattern rather than patching only the observed symptom.
\emph{Maintainability} captures preferences for proposals that fit repository constraints, reused existing abstractions, avoided unnecessary complexity, or left the system easier to extend.
\emph{Others} captures lower-frequency but still decision-grounded rationales, including choices driven by product behavior, user-facing expectations, release workflow, or delivery-speed tie-breaking when those factors determined which technically plausible proposal was preferable.
Together, these categories show that maintainer choices are not arbitrary preferences; they repeatedly appeal to reusable engineering considerations, even though the relative importance of those considerations is decided by the issue context.

\input{input/Table_proposal_selection_rationales}

\textbf{\textit{Finding 2: Proposal selection is comparative and context-conditioned.}}
By \emph{comparative decision}, we mean that maintainers do not judge each proposal only by whether it is acceptable in isolation.
Instead, they compare the salient features of candidate proposals, such as fix scope, risk, root-cause coverage, repository fit, and product behavior, and then select the proposal whose strengths best match the current issue context.
For example, a maintainer may prefer proposal A because it preserves the existing public interface, whereas proposal B requires a backward-incompatible change.
This rationale explicitly compares the candidates under a shared compatibility constraint and explains how their differences inform the selection.
All 339 sampled cases satisfy this comparative-decision pattern: each case contains multiple complete candidate proposals, and the selection evidence identifies which proposal should guide the repair under the issue context.
This supports our claim that proposal selection is made over a candidate set, where issue-specific constraints and proposal differences determine which criteria matter most.

\textbf{\textit{Finding 3: Proposal discussion helps formulate the final implementation direction.}}
By \emph{refinement evidence}, we mean discussion evidence showing that the final implementation direction is shaped through comparison, clarification, and revision around the maintainer decision, rather than copied directly from the selected proposal.
Before the decision, maintainers and proposal authors often compare candidate proposals, clarify assumptions, and surface constraints that determine which direction is most suitable.
In some threads, proposal authors explicitly contrast their own proposals with competing ones, explaining why their design better fits the issue constraints or where alternatives may be risky or incomplete; such exchanges provide maintainers with useful cues for selection.
After the decision, they may continue to refine that direction by narrowing the fix scope, rejecting risky steps, or incorporating useful observations from competing proposals.
In 273 of the 339 sampled cases (80.5\%), the issue thread contains such refinement evidence before the final implementation direction is settled.
This suggests that proposal selection is not merely an index-selection step; the surrounding discussion and comparison process both informs the maintainer's choice and helps formulate the final proposal by turning a promising direction into an implementation-facing plan.

\textbf{\textit{Implication for SWE-Manager.}}
These observations lead to one modeling hypothesis and two design requirements for SWE-Manager.
The hypothesis is that proposal selection is learnable: because maintainer choices repeatedly rely on recurring high-level rationales, the decision process is not purely ad hoc.
The first design requirement is joint comparison.
Because maintainer decisions are comparative, SWE-Manager should consider the issue and all candidate proposals together rather than score proposals in isolation.
The second design requirement is implementation-facing synthesis.
Because proposal discussions both inform selection and help formulate the final implementation direction, SWE-Manager should not only select a proposal but also explain the decision and synthesize a golden proposal that turns the selected direction and useful cross-proposal observations into actionable guidance.

%% file: input/Table_proposal_selection_rationales.tex
\begin{table*}[t]
\centering
\caption{Decision rationales for proposal selection summarized from our manual study.}
\label{tab:proposal_selection_rationales}
\footnotesize
\setlength{\tabcolsep}{4pt}
\renewcommand{\arraystretch}{1.0}
\begin{tabularx}{0.72\textwidth}{@{}>{\raggedright\arraybackslash}p{0.14\textwidth} X >{\raggedleft\arraybackslash}p{0.05\textwidth} >{\raggedleft\arraybackslash}p{0.06\textwidth}@{}}
\toprule
\textbf{Theme} & \textbf{Primary selection rationale} & \textbf{Count} & \textbf{Share} \\
\midrule
\textbf{Risk and Safety}
& Minimal, verifiable change with low regression risk & 86 & 25.4\% \\
& Controlled fix scope and high executability & 32 & 9.4\% \\
& Better robustness in edge cases & 20 & 5.9\% \\
& Time-critical mitigation & 7 & 2.1\% \\
\cmidrule(lr){2-4}
& \textit{Subtotal} & \textit{145} & \textit{42.8\%} \\
\midrule
\textbf{Fix Depth}
& Root-cause elimination (vs.\ symptom-level patching) & 68 & 20.1\% \\
& Systematic remediation (vs.\ fixing only the observed failure) & 31 & 9.1\% \\
\cmidrule(lr){2-4}
& \textit{Subtotal} & \textit{99} & \textit{29.2\%} \\
\midrule
\textbf{Maintainability}
& Conforms to repository engineering constraints & 42 & 12.4\% \\
& More extensible for future requirements or variants & 13 & 3.8\% \\
\cmidrule(lr){2-4}
& \textit{Subtotal} & \textit{55} & \textit{16.2\%} \\
\midrule
\textbf{Others}
& Better matches product design and expectations & 26 & 7.7\% \\
& Tie-breaking by delivery speed & 14 & 4.1\% \\
\cmidrule(lr){2-4}
& \textit{Subtotal} & \textit{40} & \textit{11.8\%} \\
\midrule
\textbf{Total} & & 339 & 100.0\% \\
\bottomrule
\end{tabularx}
\end{table*}

%% file: sec/004_Method.tex
\section{METHOD}
\subsection{SWE-Manager}
The manual study leads to a testable hypothesis: because maintainer choices are guided by recurring high-level rationales, proposal selection can be learned by an LLM from large-scale selection data.
It also yields two task-design requirements: the model should compare the issue and all candidate proposals jointly, and it should synthesize a golden proposal based on the selected direction while absorbing useful observations from competing proposals.
We test this hypothesis by operationalizing proposal selection as a structured generation task: given an issue and multiple candidate proposals, the model compares candidates jointly, selects the best proposal, explains the decision, and synthesizes a golden proposal.
This design mirrors the role of a technical manager in practice, where multiple proposed solutions are reviewed against feasibility, risk, maintainability, and project goals before one direction is chosen for implementation.
Our curated dataset provides evidence of such decisions by pairing open-source issues with multiple candidate proposals, a maintainer-selected best-proposal ID, and the maintainer's stated reason for that selection. Building on this dataset, we introduce {\modelName}, a reasoning-based model that selects the most suitable proposal from multiple candidates, trained via supervised fine-tuning and rule-guided reinforcement learning (RL).
We describe the methodology in detail below.

\input{input/P2A}
\subsubsection{Task Formulation}
Given a software engineering issue $I$ described in the issue thread and a set of $N$ candidate proposals $P=\{p_1,p_2,\dots,p_N\}$, {SWE-Manager} is tasked with selecting the best proposal and synthesizing a golden proposal. Specifically, the model generates a token sequence that contains: (1) a structured reasoning trace that compares candidates and key trade-offs, (2) the \emph{ID} of the selected best proposal $\hat{p}\in P$, (3) a concise explanation that justifies the selection based on evidence from the issue context, and (4) a \emph{golden proposal} $\tilde{p}$ that consolidates the strengths of the candidates into a refined plan.
Requiring \swemanager{} to synthesize a \emph{golden proposal} is intended to strengthen proposal comparison rather than to merely produce an additional artifact. The manual study suggests that maintainer decisions are not just about picking an index, but about reconciling trade-offs among safety, fix depth, repository fit, and implementation practicality. By constructing a consolidated plan that selectively integrates the strongest components from competing candidates, the model is forced to contrast proposals along their key differences, surface trade-offs, and reconcile inconsistencies. This process operationalizes the comparative reasoning observed in the study and, in turn, supports more accurate identification of the best proposal than directly predicting an index.

\subsubsection{Data Preparation}
\label{subsec:data_prep}
We use the proposal-selection corpus described in Section~\ref{subsec:data_collection}. Each instance contains the issue context, at least two structured candidate proposals, the maintainer's stated reason for choosing the selected proposal, and a ground-truth proposal ID traced from the final bounty recipient back to that contributor's proposal.
Although these records identify which proposal was selected and why, the intermediate discussion process is often fragmented and dispersed across issue-thread exchanges, making it difficult to extract as structured supervision.
Maintainers and proposal authors may compare alternatives in scattered comments, and the final implementation-facing plan is frequently absorbed directly into code rather than written as a standalone proposal.
We therefore use GPT-5 only to organize these fragmented but human-grounded records into stable training targets.
First, given the issue, the candidate proposals, the selected proposal, and the maintainer's selection reason, GPT-5 reconstructs a comparative \texttt{think} process that explains how the selected proposal differs from the alternatives under the issue context.
Second, given the candidate proposals, the selected proposal, and the maintainer's selection reason, GPT-5 synthesizes a \emph{golden proposal} by using the selected proposal as the base and incorporating useful strengths from competing proposals according to the maintainer's selection reason.
In this setup, GPT-5 does not determine the selection label; the label remains anchored to the bounty-backed maintainer decision.
Its role is to provide consistent supervision for the otherwise hard-to-observe comparison and refinement process, allowing the larger dataset to test whether proposal selection and golden-proposal synthesis can be learned at scale.
\subsubsection{Training SWE-Manager}
We train \modelName{} from Qwen3-8B in two stages, using the same structured input and output format defined above.
This recipe follows common LLM alignment practice that initializes a model with supervised fine-tuning (SFT) before reinforcement learning~\cite{Bai2022ConstitutionalAH,Ouyang2022TrainingLM,DeepSeekAI2025DeepSeekR1IR}.
First, SFT teaches the model to follow the manager-task schema: read the issue and candidate proposals, compare candidates, output the selected proposal ID, justify the choice, and synthesize a golden proposal.
Second, we optimize the model with DAPO (Decoupled Clip and Dynamic sAmpling Policy Optimization)~\cite{Yu2025DAPOAO} so that proposal selection is learned as a decision policy rather than only as imitation of model-written rationales.
This two-stage recipe separates structure learning from policy optimization: SFT provides the response format and comparative-reasoning scaffold, while reinforcement learning tests whether the model can improve proposal-selection decisions from outcome-aligned feedback.
For reinforcement learning, we use a compact weighted reward over the four structured fields: $r=0.4r_{\text{sel}}+0.2r_{\text{think}}+0.2r_{\text{justi}}+0.2r_{\text{gold}}$, where $r_{\text{sel}}=\mathbb{I}[\hat{p}=p^{*}]$ verifies whether the selected proposal matches the ground-truth proposal, and $r_{\text{think}}, r_{\text{justi}}, r_{\text{gold}}\in[0,1]$ are normalized cosine-similarity scores~\cite{Salton1975AVS} for the \texttt{think} content, the \texttt{justification}, and the synthesized golden proposal.
We implement training with MS-Swift~\cite{zhao2024swiftascalablelightweightinfrastructure} and vLLM~\cite{kwon2023efficient}, and distribute training across 12 A100 GPUs using ZeRO-3~\cite{Rajbhandari2019ZeROMO}.
In SFT, we use full-parameter fine-tuning for 200 steps with batch size 48, warmup ratio 0.1, and learning rate $5\times 10^{-6}$.
In the DAPO stage, we keep the same batch size, train for 3 epochs, set the sampling temperature to 1.0, follow the default DAPO setting of a 0.28 clip ratio without an explicit KL loss, and use learning rate $1\times 10^{-6}$.

\subsection{\frameworkName}
\label{subsec:p2a}

The SWE-Lancer Manager benchmark evaluates whether a model can select the best proposal among candidates, but it does not test whether the model's synthesized \emph{golden proposal} can guide downstream implementation.
To evaluate this second capability, we introduce \frameworkName{}, a proposal-first, implementation-later workflow.
\frameworkName{} first generates candidate proposals, then asks \swemanager{} to select and synthesize a golden proposal, and finally uses that golden proposal to guide patch generation.

As shown in Fig.~\ref{fig:P2A}, \frameworkName{} contains three roles.
The \proposalagent{} reads the issue context, inspects the repository, and generates candidate proposals, each containing a problem restatement, a root-cause analysis, and a proposed solution.
The \technicalmanager{} compares the candidate proposals under the issue constraints, outputs the selected proposal, explains the selection, and synthesizes a \emph{golden proposal}.
The \implementationagent{} then follows the golden proposal to implement the required code changes and produce the final patch.
Both the \proposalagent{} and the \implementationagent{} are adapted from mini-SWE-agent~\cite{yang2024sweagent}, a lightweight bash-driven software engineering agent that interacts with the repository through shell commands.

In the main \frameworkName{} workflow, we choose the synthesized \emph{golden proposal} as the artifact passed to the \implementationagent{}.
This choice matches the capability that \frameworkName{} is designed to evaluate: whether \swemanager{}'s proposal synthesis can provide useful guidance for downstream issue resolution, beyond predicting the selected proposal ID.
The choice also reflects a key difference between the two benchmarks.
In the SWE-Lancer Manager benchmark, one candidate is guaranteed to correspond to the eventual ground-truth solution, whereas the proposals generated on the fly by the \proposalagent{} may all be incorrect or incomplete.
Forcing the \implementationagent{} to follow one of them would then likely yield an incorrect patch, because downstream execution inherits the flaws of the chosen proposal; even if the \technicalmanager{} recognizes that every candidate is inadequate, selection alone gives it no channel to act on this judgment.
The golden proposal provides this channel: the \technicalmanager{} can cross-analyze the candidates, keep their correct elements, and revise or discard flawed steps before implementation begins.
In the ablation study (RQ3, Section~\ref{subsec:rq3}), we replace the golden proposal with the selected proposal to test whether synthesis provides additional benefit over selection alone.
By holding the proposal and implementation agents fixed, \frameworkName{} uses the final issue-resolution result as a proxy for the downstream usefulness of the synthesized golden proposal.

%% file: input/P2A.tex
\begin{figure*}[t]
  \centering
  \includegraphics[width=\textwidth]{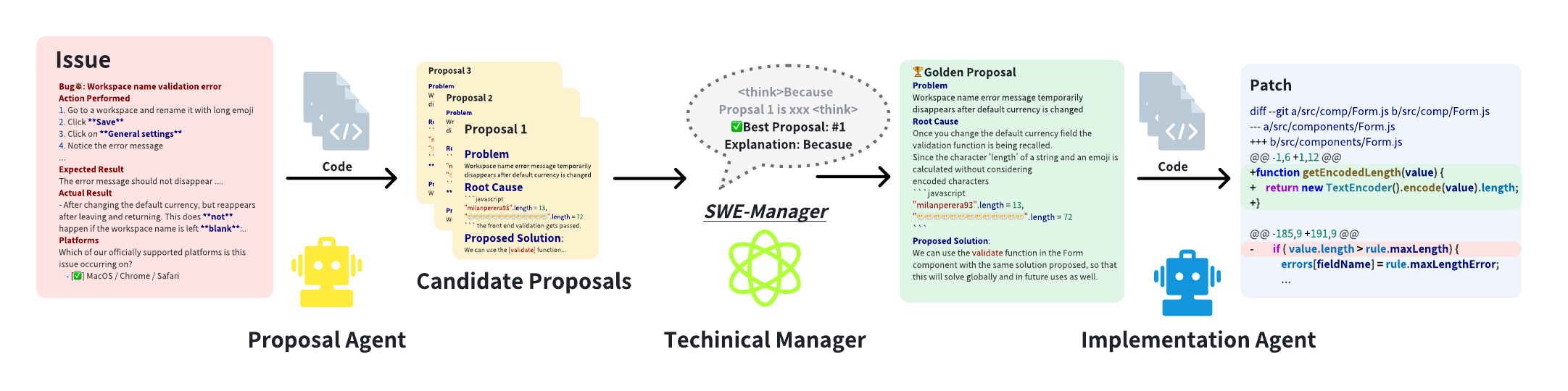}
  \Description{A three-stage workflow connects proposal agents, a technical manager, and an implementation agent. Candidate proposals are compared and synthesized into a final plan, which guides generation of a code patch.}
  \caption{\textbf{P2A} comprises three roles: \textbf{Proposal Agent}, \textbf{Technical Manager}, and \textbf{Implementation Agent}. The Proposal Agent samples multiple candidate proposals; each proposal includes an issue restatement, a root-cause analysis, and a proposed solution. The Technical Manager compares proposals and decides on the final proposal for implementation. \textbf{SWE-Manager plays a Technical Manager role in P2A.} Finally, the Implementation Agent follows the golden proposal to produce a code patch that resolves the issue.}

  \label{fig:P2A}
\end{figure*}

%% file: sec/005_Experiment.tex
\section{EXPERIMENT}
In this section, we first test the central hypothesis of the paper: whether proposal selection is learnable by a small model despite appearing highly contextual and case-specific. We then validate whether this learned proposal-reasoning capability is useful in a realistic workflow, testing whether proposal reasoning and selection prior to coding improves end-to-end issue resolution. Finally, ablations and analyses identify which task and training choices are most responsible for making this capability learnable in practice.

\subsection{Benchmarks}
\label{subsec:setup}

We evaluate \swemanager{} on the SWE-Lancer benchmark suite~\cite{Miserendino2025SWELancerCF} released by OpenAI, which comprises two benchmarks: Manager and IC.
SWE-Lancer Manager evaluates proposal selection: given an issue description and multiple candidate proposals, the goal is to select the best proposal.
It contains 265 instances with a total reward budget of \$264{,}500.
Each instance includes a user-reported issue and structured candidate proposals.
The issue typically reports reproduction steps, observed and expected behavior, and environment details, while each proposal provides a problem restatement, root-cause analysis, and fix strategy.
\swemanager{} is further validated on SWE-Lancer IC within the \frameworkName{} framework to assess whether the synthesized golden proposal improves downstream implementation. SWE-Lancer IC measures end-to-end issue resolution via test-based verification of the generated patch and contains 198 instances with a total reward budget of \$189{,}300.

\subsection{Baselines}
\label{subsec:baselines}

\subsubsection{SWE-Lancer Manager benchmark baselines - \textbf{Direct Selection} and \textbf{Agent Selection}}
For the SWE-Lancer Manager benchmark, we evaluate two proposal-selection settings.
In the \textbf{direct selection} setting, the model receives the issue description and candidate proposals, then directly outputs the selected proposal.
This setting does not allow codebase access and is instantiated with GPT-5, Qwen3-8B, and \modelName{}.
In the \textbf{agent selection} setting, the selector uses the same issue and proposal input and produces the same selected-proposal output, but can additionally inspect the repository code before making the selection, following the code-access setting in SWE-Lancer Agent~\cite{Miserendino2025SWELancerCF}.
Thus, the only difference between the two settings is whether the selector can access code.
In Table~\ref{tab:rq1_manager}, we mark the code-access setting as \emph{with Agent}; rows without this suffix use the direct selection setting.

\subsubsection{SWE-Lancer IC benchmark baselines - \textbf{mini-SWE-agent} and \textbf{P2A with different model}}
For the SWE-Lancer IC benchmark, we compare \frameworkName{} with baselines that separate direct implementation ability from the effect of the Technical Manager.
\textbf{mini-SWE-agent} is the direct implementation baseline: given an issue, it attempts to generate a patch without an explicit proposal-selection and synthesis stage.
We instantiate mini-SWE-agent with different LLM backbones to provide a strong reference for execution-based patch success.
Because the \proposalagent and \implementationagent in \frameworkName{} are also based on mini-SWE-agent, this comparison isolates the additional value of inserting the Technical Manager between proposal generation and implementation.
Within \frameworkName{}, we further evaluate two manager-replacement baselines that use GPT-5 or the pre-training backbone Qwen3-8B instead of \modelName{}.
These variants keep the same proposal agents, implementation agent, and candidate proposal pool, thereby isolating the effect of the learned manager model.

\subsection{Metrics}
\label{subsec:metrics}

\subsubsection{SWE-Lancer Manager (\textbf{Match (\%), \#Match, Earned (\%), \$Earned})}
We evaluate proposal selection by checking whether the model-selected proposal matches the ground-truth proposal for each issue. We report {Match (\%)}, the percentage of issues for which the selected proposal ID matches the ground-truth proposal ID. We additionally report {\#Match}, the absolute number of issues for which the selected proposal matches the ground truth. Following SWE-Lancer's scoring scheme, we report {Earned (\%)}, the percentage of the total available reward earned by the model on the test set, and {\$Earned}, the total dollar amount earned by correct selections.

\subsubsection{SWE-Lancer IC (\textbf{Pass (\%), \#Pass, Earned (\%), \$Earned})}
We evaluate downstream issue resolution using the benchmark's execution-based criterion. We report {Pass (\%)}, defined as the percentage of issues for which the produced code patch passes the provided verification and resolves the issue. We additionally report {\#Pass}, the absolute number of issues successfully resolved. We also report {Earned (\%)}, defined as the percentage of the total available reward that is earned by successful patches on the test set, and {\$Earned}, defined as the total dollar amount earned by successful patches.

\subsection{\textbf{RQ1:} How well does SWE-Manager learn to select the best proposal?}

We answer this question by comparing a trained 8B model, \modelName{}, against the pre-training backbone model Qwen3-8B and GPT-5 in the direct selection setting, and GPT-5 with code-access agent selection on the SWE-Lancer Manager benchmark, using the ground-truth match rate {Match (\%)} and the earned rate {Earned (\%)} as the primary metrics. This comparison directly tests whether proposal selection can be systematically learned by a small model. As shown in Table~\ref{tab:rq1_manager}, {\modelName{} achieves the best overall performance, reaching {53.21\%} match rate, {57.75\%} earned rate, and earning \$152,750}. It surpasses GPT-5 and Qwen3-8B in the direct selection setting, and also outperforms GPT-5 with Agent, which is able to access the codebase during selection.

\input{input/Table_RQ1}

\input{input/price_and_proposals_distribution}
To better understand where the gains come from, Fig.~\ref{fig:comb_distri} breaks down match rate by (i) the number of competing proposals and (ii) the instance reward level.
We first analyze performance by proposal-count distribution. As the number of candidate proposals increases, the selection accuracy generally decreases across all methods, reflecting the growing difficulty of comparative judgment in a larger proposal space. Nevertheless, \modelName{} consistently outperforms the direct selection baselines, GPT-5 and Qwen3-8B, across all bins. For instances with only two proposals, \modelName{} achieves the best accuracy of 74.36\% (29/39). For instances with three proposals, \modelName{} reaches 62.30\% (38/61) and outperforms direct GPT-5 by 18.04\% and Qwen3-8B by 32.79\%, demonstrating the biggest performance gap. For larger candidate proposal sets, \modelName{} remains competitive and continues to outperform the direct baselines, achieving 47.22\% (17/36) for four proposals, 50.88\% (29/57) for five proposals, and 38.89\% (28/72) for six or more proposals.
These results suggest that decision-aligned RL training improves comparative proposal judgment in realistic multi-proposal scenarios.

The reward-stratified view further highlights a distinction between direct selection and the code-access agent setting. \modelName{} leads in the low-to-mid reward ranges, reaching 45.28\% at \$0--\$500, 58.59\% at \$500--\$1K, and 54.35\% at \$1K--\$2K. However, for the highest-value instances (\$2K+), GPT-5 with Agent achieves the best accuracy at 52.38\%, outperforming \modelName{}. A similar pattern appears in the most crowded proposal sets, where GPT-5 with Agent slightly surpasses \modelName{} at six or more proposals. These results indicate that code access is not a universal remedy for proposal selection, but can be particularly beneficial for the most complex or highest reward cases.

\textbf{Answer to RQ1.} The proposal-selection policy is learnable by a small model. With only 8B parameters, \modelName{} improves over its Qwen3-8B backbone from 34.72\% to 53.21\% in Match (\%) and from 39.98\% to 57.75\% in Earned (\%), while also surpassing direct GPT-5 by 9.06 percentage points in Match (\%) and 20.00 percentage points in Earned (\%). Its gains across most proposal-count groups and reward tiers suggest that task-specific training strengthens comparative proposal selection rather than only improving a narrow subset of cases.

\subsection{\textbf{RQ2:} Does the synthesized golden proposal improve downstream issue resolution?}
Having established that proposal selection is learnable, we use RQ2 to evaluate the second capability defined in our task formulation: whether a synthesized golden proposal can guide downstream implementation.
We compare two settings on SWE-Lancer IC.
First, mini-SWE-agent directly generates patches from the issue without an explicit proposal-selection or synthesis stage.
Second, in the main P2A setting, proposal agents generate candidate proposals, a Technical Manager selects and synthesizes a golden proposal, and a fixed Implementation Agent follows the golden proposal to produce the patch.
We defer the comparison between the golden proposal and the selected proposal to the ablation study in RQ3.
\input{input/Table_RQ2}
The upper part of Table~\ref{tab:rq2_ic} reports the direct implementation baselines. The strongest baseline by pass rate resolves 96 of 198 issues with a 48.5\% pass rate.
We use these direct-baseline results to configure P2A.
Because proposal generation and patch generation require repeated agent executions, we do not use GPT-5 as either a proposal source or the implementation model in P2A.
To build a meaningful diverse proposal-selection pool, we generate candidate proposals with multiple models rather than relying on a single proposal generator.
Specifically, the proposal sources are the top three models by pass rate in the upper part of Table~\ref{tab:rq2_ic}: Gemini 3 Flash, DeepSeek V3.2, and GLM-4.7.
We then fix the Implementation Agent to Gemini 3 Flash, the strongest direct implementation baseline, across all P2A variants, and instantiate the Technical Manager with GPT-5, Qwen3-8B, and \modelName{}.

The lower part of Table~\ref{tab:rq2_ic} reports the downstream results when the Implementation Agent follows the golden proposal synthesized by each Technical Manager.
With \modelName{} as the Technical Manager, P2A resolves 110 of 198 issues, reaching a 55.6\% pass rate and improving over the strongest direct implementation baseline by 7.1 percentage points.
The comparison with Qwen3-8B further shows the effect of learning the manager policy, increasing resolved issues from 97 to 110 and raising the pass rate from 49.0\% to 55.6\%.
\modelName{} also matches the pass rate of P2A with GPT-5, although GPT-5 earns more reward.

To understand where these gains come from, Fig.~\ref{fig:rq2_venn} compares the sets of instances solved by different systems.
Panels (a) and (b) compare P2A with the strongest direct baseline, mini-SWE-agent with Gemini 3 Flash, which uses the same model as the Implementation Agent in P2A.
With \modelName{} as the Technical Manager, P2A solves 31 instances that the baseline does not, which account for 28.2\% of its 110 successes; with GPT-5, it solves 35 such instances.
Because the Implementation Agent and the baseline use the same model, these additional successes come from the guidance of the golden proposal rather than from a stronger implementation model.

\input{input/venn}

Panel (c) highlights the main benefit of the Technical Manager.
It compares P2A with \modelName{} against the \emph{Union} of instances solved by any of the three proposal-source models under mini-SWE-agent.
P2A with \modelName{} solves 9 instances outside the Union, that is, issues that none of the three proposal-source models resolves on its own.
In these cases, every candidate proposal comes from a model that cannot resolve the issue by itself, yet the golden proposal synthesized by \modelName{} leads to a successful patch.
This shows that \modelName{} does more than pick the best of its inputs: by consolidating complementary partial insights from imperfect proposals, it enables the Implementation Agent to resolve issues that are beyond the reach of every individual proposal source.

Panel (d) compares \modelName{} with GPT-5 as the Technical Manager under the same proposal pool.
They share 91 solved instances, and each solves 19 instances that the other does not.
Although the supervision targets of \modelName{} are generated by GPT-5 (Section~\ref{subsec:data_prep}), the model therefore does not simply imitate GPT-5.
One explanation is that its selection labels come from maintainer decisions rather than from GPT-5, and reinforcement learning further optimizes the model against these labels.
The partly different solved subsets also explain why the two Technical Managers reach the same pass rate but different earned rates in Table~\ref{tab:rq2_ic}, because rewards vary across instances.

\textbf{Answer to RQ2.} \frameworkName{} with \modelName{} achieves a 55.6\% pass rate, matching P2A with GPT-5 and improving over the strongest mini-SWE-agent baseline by 7.1\%. Its additional successes are largely distinct from those of the direct baseline and include 9 issues that none of the three proposal-source models resolves alone, showing that the synthesized golden proposal can guide the Implementation Agent beyond the reach of every individual proposal source. Its solved issues also differ in part from those of P2A with GPT-5, indicating that \modelName{} does not merely imitate its GPT-5-generated supervision.

\subsection{\textbf{RQ3:} Which design choices drive SWE-Manager's gains?}
\label{subsec:rq3}
Beyond reporting SWE-Lancer Manager benchmark scores, we conduct an ablation study to isolate which design and training choices are responsible for \modelName{}'s improvement. In our main setting, the model is trained to both select the best proposal and synthesize a golden proposal that consolidates complementary strengths across candidates. To test whether this synthesis objective is meaningful for selection quality, we construct a matched training set and train a variant that is asked to \emph{only} predict the best proposal ID and reason. This allows us to ask not just whether the model improves, but which task formulation and training recipe make proposal selection learnable in practice. We further ablate the proposal artifact that P2A passes to the Implementation Agent, to test whether synthesis, rather than selection alone, is what makes the Technical Manager useful downstream.

\input{input/swe-manager_ablation_study}

In Fig.~\ref{fig:rq3_ablation}, the yellow curve corresponds to the variant trained to \emph{only} predict the best proposal ID on the SWE-Lancer Manager benchmark. From the beginning, this variant consistently underperforms the RL-trained alternatives in both ID accuracy and earned reward, and the gap persists throughout training. Although its performance increases slowly as training proceeds, it remains well below the configurations that optimize a richer objective. This pattern suggests that index-only supervision provides limited learning signal: the model can memorize shallow correlations for choosing an ID, but it is less effective at learning the comparative reasoning needed to distinguish among multiple plausible proposals. In contrast, requiring the model to synthesize a golden proposal encourages a more explicit evaluation of candidate strengths and weaknesses, which translates into higher selection accuracy and reward.

We further ablate the training recipe by comparing {SFT+DAPO} against applying DAPO directly. In Fig.~\ref{fig:rq3_ablation}, the green curve denotes the model initialized with supervised fine-tuning 200 steps before reinforcement learning, whereas the blue curve denotes training with DAPO from the base model without SFT. Overall, the green curve is consistently stronger and more stable across training, reaching the best final accuracy of 53.21\% and an earned rate of 57.75\%. This suggests that SFT provides a useful warm start by teaching the model the desired reasoning structure and output format, thereby making subsequent policy optimization more effective. By contrast, directly applying DAPO appears more sensitive to early optimization dynamics and converges to a weaker solution, indicating that learning the reasoning scaffold and learning the selection policy are best staged rather than attempted solely through RL from scratch.

\textbf{Golden versus selected proposal in P2A.}
As discussed in Section~\ref{subsec:p2a}, the candidate proposals in P2A are generated on the fly, so unlike in the SWE-Lancer Manager benchmark, no candidate is guaranteed to correspond to a correct solution.
Following a single selected proposal therefore inherits its flaws, whereas the golden proposal allows the Technical Manager to revise them.
To test this design choice, we keep the proposal pool and the Implementation Agent fixed and replace the golden proposal of \modelName{} with the original proposal it selected.
Table~\ref{tab:rq3_golden_selected} reports the results.

\input{input/Table_RQ3_golden_selected}

The golden proposal resolves 18 more issues than the selected proposal, raising the pass rate from 46.5\% to 55.6\% and the earned reward from \$58{,}125 to \$69{,}875.

The weaker result of the selected proposal is consistent with how the model is trained.
Its supervision uses the selected proposal as the base of the golden proposal, which then incorporates and corrects observations from the competing proposals (Section~\ref{subsec:data_prep}).
The selected proposal is thus the starting point of the model's plan rather than a plan intended to be executed verbatim.
Moreover, selection accuracy on the SWE-Lancer Manager benchmark, where a correct candidate is guaranteed, does not directly transfer to P2A, where every candidate may be flawed.
What matters downstream is whether the Technical Manager can turn the candidates into a better plan, which is exactly what the golden proposal provides.

\textbf{Answer to RQ3.} The ablation study shows that proposal selection improves when training requires golden-proposal synthesis rather than only best-proposal ID prediction, and that supervised fine-tuning provides an effective initialization for subsequent DAPO optimization. Together, these results indicate that the gains of \modelName{} arise from both a richer selection-and-synthesis task formulation and a staged training path that first teaches the reasoning format before applying reinforcement learning. In P2A, passing the golden proposal instead of the selected proposal resolves 18 more issues with \modelName{} as the Technical Manager, which confirms that synthesis, rather than selection alone, makes the Technical Manager useful downstream.

%% file: input/Table_RQ1.tex
\begin{table}[htbp]
\centering
\caption{Performance on the SWE-Lancer Manager benchmark.}
\label{tab:rq1_manager}
\normalsize
\setlength{\tabcolsep}{6pt}
\renewcommand{\arraystretch}{1.15}
\begin{tabular*}{0.95\linewidth}{@{}l@{\extracolsep{\fill}}rrrr@{}}
\toprule
\textbf{System} & \textbf{Match (\%)} & \textbf{\#Match} & \textbf{Earned (\%)} & \textbf{\$Earned} \\
\midrule
GPT-5 with Agent     & 47.20 & 125 & 54.35 & \$143,750 \\
GPT-5                & 44.15 & 117 & 37.75 & \$99,875 \\
Qwen3-8B             & 34.72 &  92 & 39.98 & \$105,750 \\
\textbf{\modelName{}} & \textbf{53.21} & \textbf{141} & \textbf{57.75} & \textbf{\$152,750} \\
\bottomrule
\end{tabular*}
\end{table}

%% file: input/price_and_proposals_distribution.tex
\begin{figure}[!htp]
  \centering
  \includegraphics[width=0.85\linewidth]{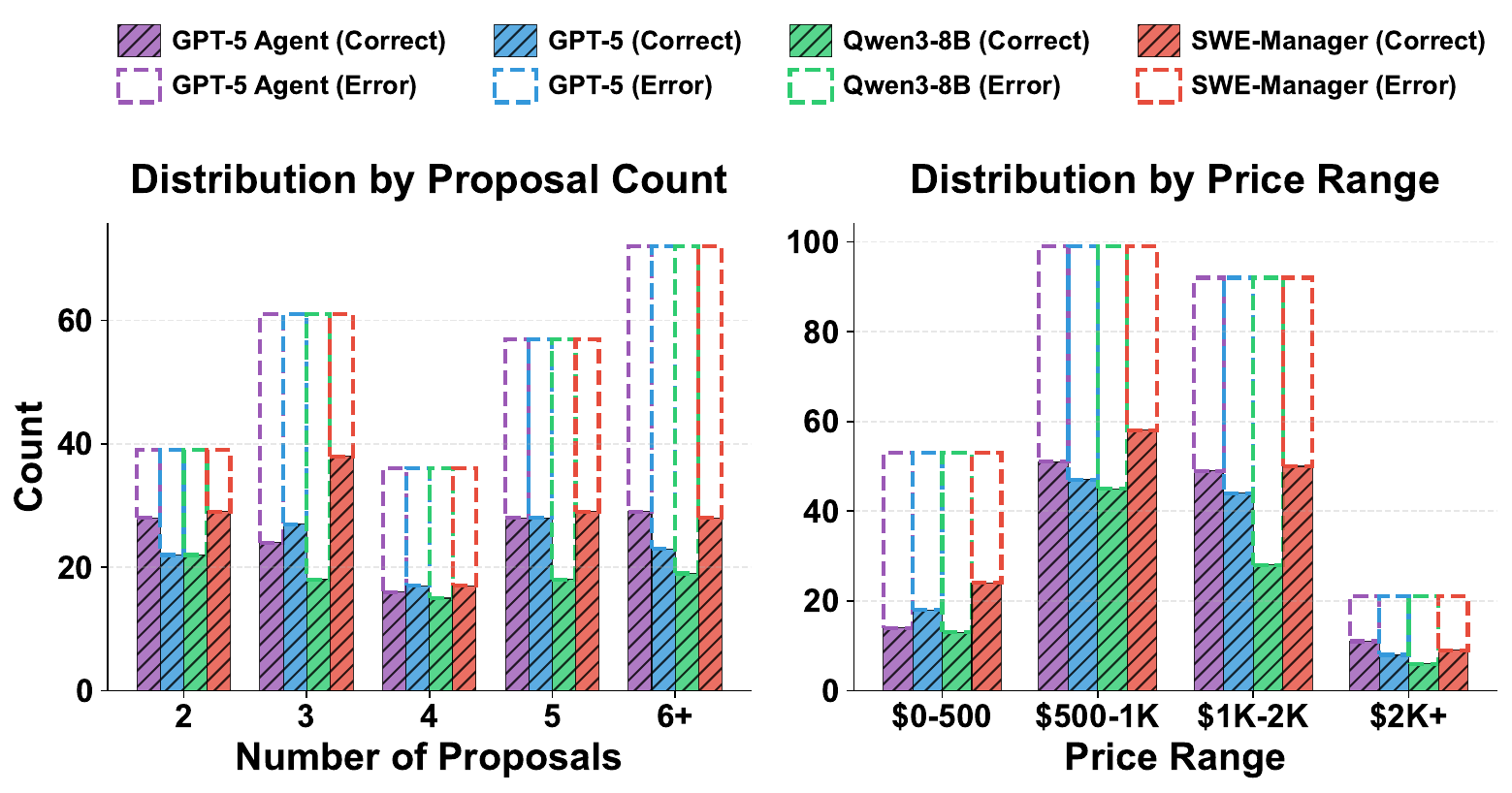}
  \Description{Two grouped bar charts show correct and incorrect proposal selections for GPT-5 with code access, direct GPT-5, Qwen3-8B, and SWE-Manager. The left chart groups issues by candidate count; the right groups issues by bounty range.}
\caption{Distribution of correct and incorrect selections on the SWE-Lancer Manager benchmark. \textbf{Left:} results grouped by the number of candidate proposals; \textbf{Right:} results grouped by task reward. Performance generally declines with larger proposal sets, while {SWE-Manager} remains strongest in most bins.}

  \label{fig:comb_distri}
\end{figure}

%% file: input/Table_RQ2.tex
\begin{table}[htbp]
\centering
\caption{Issue resolution on the SWE-Lancer IC benchmark. \textbf{Upper:} mini-SWE-agent implements patches directly with different backbone models. \textbf{Lower:} P2A with different Technical Managers; candidate proposals come from Gemini 3 Flash, DeepSeek V3.2, and GLM-4.7, and the Implementation Agent (Gemini 3 Flash) follows the golden proposal.}
\label{tab:rq2_ic}
\normalsize
\setlength{\tabcolsep}{6pt}
\renewcommand{\arraystretch}{1.15}
\begin{tabular*}{0.95\linewidth}{@{}l@{\extracolsep{\fill}}rrrr@{}}
\toprule
\textbf{System} & \textbf{Pass (\%)} & \textbf{\#Pass} & \textbf{Earned (\%)} & \textbf{\$Earned} \\
\midrule
\multicolumn{5}{@{}l}{\textit{mini-SWE-agent (backbone model)}} \\
GPT-5-mini       & 34.8 & 69 & 24.1 & \$45{,}625 \\
Claude Haiku 4.5 & 42.9 & 85 & 42.5 & \$80{,}425 \\
GLM-4.7          & 44.4 & 88 & 34.0 & \$64{,}375 \\
DeepSeek V3.2    & 46.0 & 91 & 37.4 & \$70{,}875 \\
Gemini 3 Flash   & 48.5 & 96 & 37.4 & \$70{,}875 \\
GPT-5            & 48.5 & 96 & 37.3 & \$70{,}625 \\
\midrule
\multicolumn{5}{@{}l}{\textit{P2A (Technical Manager)}} \\
GPT-5                 & \textbf{55.6} & \textbf{110} & \textbf{45.8} & \textbf{\$86{,}625} \\
Qwen3-8B              & 49.0 & 97 & 33.5 & \$63{,}375 \\
\textbf{\modelName{}} & \textbf{55.6} & \textbf{110} & 36.9 & \$69{,}875 \\
\bottomrule
\end{tabular*}
\end{table}

%% file: input/venn.tex
\begin{figure}[!htp]
  \centering
  \includegraphics[width=\linewidth]{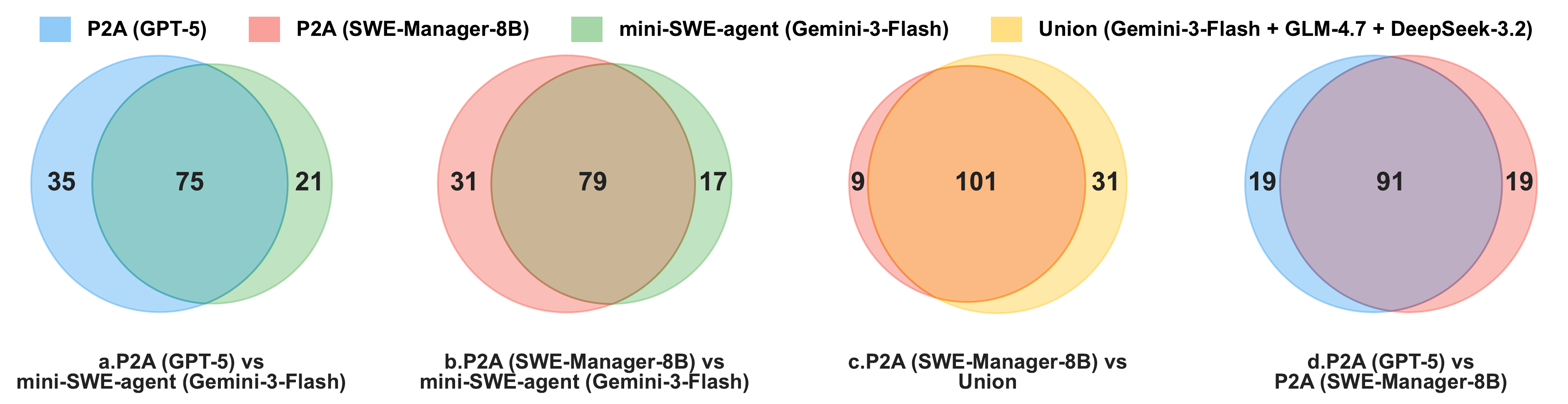}
  \Description{Four Venn diagrams compare the SWE-Lancer IC instances solved by P2A with GPT-5, P2A with SWE-Manager-8B, mini-SWE-agent with Gemini 3 Flash, and the union of the three proposal-source models.}
  \caption{Overlap of solved SWE-Lancer IC instances. (a) and (b): P2A with GPT-5 or \modelName{} as the Technical Manager versus mini-SWE-agent with Gemini 3 Flash. (c): P2A with \modelName{} versus the \emph{Union} of instances solved by any of the three proposal-source models (Gemini 3 Flash, GLM-4.7, and DeepSeek V3.2) under mini-SWE-agent. (d): P2A with GPT-5 versus P2A with \modelName{}.}
  \label{fig:rq2_venn}
\end{figure}

%% file: input/swe-manager_ablation_study.tex
\begin{figure}[!htp]
  \centering
  \includegraphics[width=0.85\linewidth]{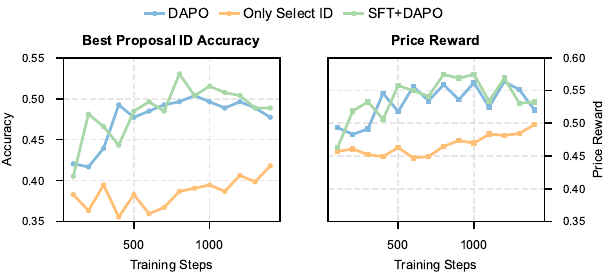}
  \Description{Two line charts plot selection accuracy and reward against training steps for DAPO, ID-focused training, and SFT followed by DAPO. The charts compare the training trajectories of these three variants.}
\caption{Training ablation of \modelName{} on the SWE-Lancer Manager benchmark. {Only Select Proposal ID} tests ID-only training without golden-proposal synthesis, while {DAPO} and {SFT+DAPO} compare direct reinforcement learning with an SFT warm start.}
  \label{fig:rq3_ablation}
\end{figure}

%% file: input/Table_RQ3_golden_selected.tex
\begin{table}[htbp]
\centering
\caption{Ablation of the proposal artifact passed to the Implementation Agent in P2A on SWE-Lancer IC, with \modelName{} as the Technical Manager.}
\label{tab:rq3_golden_selected}
\normalsize
\setlength{\tabcolsep}{6pt}
\renewcommand{\arraystretch}{1.15}
\begin{tabular*}{0.95\linewidth}{@{}l@{\extracolsep{\fill}}rrrr@{}}
\toprule
\textbf{Proposal artifact} & \textbf{Pass (\%)} & \textbf{\#Pass} & \textbf{Earned (\%)} & \textbf{\$Earned} \\
\midrule
\textbf{Golden proposal} & \textbf{55.6} & \textbf{110} & \textbf{36.9} & \textbf{\$69{,}875} \\
Selected proposal        & 46.5 & 92 & 30.7 & \$58{,}125 \\
\bottomrule
\end{tabular*}
\end{table}

%% file: sec/006_Conclusion.tex
\section{DISCUSSION}

\subsection{Robustness Under Larger Proposal Pools}
Proposal selection becomes harder as the candidate pool grows, so we further examine whether the learned selector remains robust when P2A supplies more proposals.
Fig.~\ref{fig:rq2_ablation} shows that \modelName{} remains close to GPT-5 when the pool contains two or three proposals, which matches the dominant regime in our training data.
The gap becomes larger once the pool expands to four or five proposals: GPT-5 remains comparatively stable, whereas \modelName{} loses more accuracy and reward.

\input{input/selector_comparison_among_proposal_num}

This degradation is consistent with a distribution shift rather than a failure of the task formulation itself.
As the pool grows, the selector must compare more alternatives and the input length increases sharply; at the same time, the training set contains relatively few large-pool examples.
Thus, four- and five-proposal cases ask \modelName{} to operate in a longer and less represented regime.
The result suggests that the current model has learned useful comparative behavior, but its robustness is still limited by training coverage and long-context comparison difficulty.

\subsection{Threats to Validity}

\subsubsection{Data Leakage}
We mitigate potential data leakage and contamination effects. When constructing the training set, we explicitly remove all instances from the SWE-Lancer Manager and SWE-Lancer IC benchmarks. Cross-issue leakage is also unlikely because candidate proposals are written in a structured natural-language format that mainly includes a problem restatement, root-cause analysis, and a proposed solution, and any concrete code they include is typically limited, partial, or illustrative rather than a full implementation. As a result, proposals from later issues are unlikely to encode implementation details that would reveal the ground-truth best proposal for earlier benchmark issues.

\subsubsection{Generalizability and Robustness}
In the P2A framework, \modelName{} functions as both selector and synthesizer for proposals generated on demand by proposal agents. Since these candidate proposals are produced at inference time rather than drawn from the training corpus, P2A evaluates selection and synthesis on genuinely unseen proposal sets, making it less likely that performance is driven by memorizing specific training instances.
At the same time, our results expose a robustness boundary for larger proposal pools. Most training instances contain only two or three proposals, whereas practical settings may present richer candidate sets. Once the pool grows beyond three proposals, the input context becomes substantially longer and \modelName{} must compare candidates in a regime that is sparsely represented during training. This distribution shift can weaken comparative judgment over longer, denser inputs, helping explain why \modelName{} falls behind GPT-5 on larger proposal sets.

\subsubsection{Conclusion Reliability}
Manual study of selection rationales may be subject to researcher bias and category ambiguity. We mitigate this threat via cross-checking and adjudication, and we provide the taxonomy definitions and representative examples for transparency. End-to-end IC evaluation is computationally expensive, and the behavior of closed-source models may change over time. We therefore document inference settings and model versions, and we release prompts and scripts when possible to facilitate replication.

\section{LIMITATIONS AND FUTURE WORK}
Our study has two main limitations, both of which suggest concrete future directions.
First, the current \swemanager{} operates without code access. Although it outperforms the GPT-5 agent baseline overall, code access may still help on high-value or repository-specific cases; a natural extension is a tool-augmented manager that can selectively inspect repository context while preserving the same decision-oriented interface.
Second, robustness remains limited when the proposal pool grows and the input context becomes substantially longer. Future work should rebalance training toward larger proposal pools, add synthesized multi-proposal examples, and use curriculum or long-context comparison strategies that gradually increase proposal count and context length.

\section{RELATED WORK}
\subsection{Proposal Selection}\label{sec:background:ps}
Proposal selection is a recurring decision problem across the software engineering life cycle.
It appears when engineers choose architectural styles, third-party components, refactoring opportunities, or candidate patches~\cite{V2018GroupDI, Lenarduzzi2021ASL}. In each setting, the core challenge is to compare multiple candidate solutions to the same software problem while jointly considering functional requirements, non-functional attributes (\eg{} performance, security, maintainability), resource constraints, and potential risks.

This decision is difficult because software design problems are often \textit{wicked problems}: candidate solutions are rarely simply correct or incorrect, but instead expose competing merits and trade-offs.
Prior work has addressed this complexity from both human-centered and automated perspectives. Rational-analysis methods such as ATAM~\cite{706657} and CBAM~\cite{Kazman2001CBAM} provide structured processes for experts to externalize and justify design trade-offs, while recent ML- and LLM-based work increasingly treats selection as a computational ranking problem, as in patch ranking for automated program repair~\cite{10.1145/3533767.3534368, xu2025scalablesupervisingsoftwareagents} and tree-search-based generation strategies~\cite{Antoniades2024SWESearchES}. \swemanager{} builds on this decision-oriented view, but focuses on learning issue-specific selection and synthesis over proposal pools before implementation begins.

\subsection{Multi-Agent Frameworks and Candidate Selection}\label{sec:related:mas}
LLM-based software engineering systems increasingly divide the work across multiple agents or stages, and several of them produce more than one candidate for the same task.
Table~\ref{tab:related_selection} contrasts these systems with \swemanager{} along four questions: what the candidates are, when the selection is made, which signal decides it, and what happens to the candidates that are not selected.

\input{input/Table_related_selection}

\textbf{Role-based frameworks.}
MetaGPT~\cite{Hong2024MetaGPT} encodes standard operating procedures into roles such as Product Manager, Architect, Project Manager, and Engineer, which work sequentially on a single requirements document and system design.
ChatDev~\cite{Qian2024ChatDev} organizes a CEO, CTO, programmer, reviewer, and tester into a chat chain in which each subtask is settled through one instructor--assistant dialogue.
MAGIS~\cite{Tao2024MAGIS} targets GitHub issue resolution: its Manager decomposes the issue into file-level tasks, assigns them to Developer agents, and produces one work plan.
Although these frameworks produce plans, they do not assess or compare the quality of the plans themselves: each plan is passed directly to implementation, and quality checks, where present, target the generated code rather than the plan.

\textbf{Plan-then-code generation.}
Self-planning~\cite{Jiang2024SelfPlanning} derives one plan and then generates code conditioned on it.
MapCoder~\cite{Islam2024MapCoder} drafts several natural-language plans before coding, orders them by the model's self-reported confidence, and accepts the first plan whose code passes the sample input/output tests, falling back to the next plan otherwise.
PlanSearch~\cite{Wang2025PlanSearch} searches over natural-language plans to diversify generation, implements each plan, and is evaluated with pass@$k$, optionally after filtering programs with public tests.
In both MapCoder and PlanSearch, the plans come from the same model and are judged by whether their programs pass tests.

\textbf{Patch-level selection for issue resolution.}
Agentless~\cite{10.1145/3715754} samples multiple patches, filters them with regression and reproduction tests, and submits the majority-voted patch.
SWE-Search~\cite{Antoniades2024SWESearchES} selects among final patches through a multi-agent debate, and trained verifiers judge agent trajectories or patches~\cite{pan2025training, xu2025scalablesupervisingsoftwareagents}.
These approaches select after implementation, so every candidate must first be turned into a complete patch, and the patches that are not selected are discarded.

\textbf{Positioning of \swemanager{}.}
Across these systems, selection, when it exists, is a verification step: tests, votes, debates, or the model's self-reported confidence establish whether a candidate works.
Such signals say little about how several workable candidates differ on criteria that tests rarely capture, such as regression risk, fix depth, and maintainability, which are the criteria maintainers weigh when several proposals could all fix the issue (Section~\ref{subsec:reasons}).
Proposal selection in the Manager task is instead a judgment problem: the decision is made before any code exists, no execution signal is available, and the reference decision comes from human maintainers rather than from a test oracle.
\swemanager{} differs from the systems above in three ways.
First, it selects among natural-language proposals before any code is written, so only the chosen direction is implemented and no execution signal is available to the selector.
Second, its selection criterion is learned from real maintainer decisions instead of being derived from tests, votes, or the model's self-assessment.
Third, it does not discard the non-selected proposals: it synthesizes a golden proposal that carries their useful observations into the implementation plan.
These differences make \swemanager{} complementary to the systems above rather than a replacement for them; for example, it could rank the plans that MapCoder drafts, or screen proposals before patch-level selection in an Agentless-style pipeline.

\subsection{Code Generation and Bug Fixing}

Large language models (LLMs) based on transformer architectures have reshaped code generation and bug fixing~\cite{10.1145/3747588}. Progress has been driven by large-scale repository pretraining and standardized datasets or benchmarks for reproducible evaluation~\cite{pan2025training,yang2025swesmith, Kocetkov2022TheStack, Jimenez2023SWEbenchCL,zhang2025swebench}, supporting both \emph{model-centric} advances that train open-source models for issue resolution~\cite{Xie2025SWEFixerTO, pan2025training} and \emph{system-centric} approaches that orchestrate LLMs for software engineering tasks, which we discuss below.
A major line of work builds \emph{agent-based} or \emph{workflow-based} systems that resolve issues end-to-end. \textit{SWE-Agent}~\cite{yang2024sweagent} lets an LLM iteratively inspect repositories, edit files, and run tests, while OpenHands~\cite{wang2025openhands} generalizes this tool-using paradigm to broader multi-step software engineering automation. Agentless~\cite{10.1145/3715754} instead improves reliability by decomposing issue resolution into controlled stages such as localization and patching.
The automated program repair (APR) community provides a complementary foundation for controlled evaluation and patch generation. Defects4J pairs real Java bugs with regression test suites for execution-based comparison~\cite{Just2014Defects4JAD}, and recent LLM-based repair systems such as D4C~\cite{10.1109/ICSE55347.2025.00169} and ChatRepair~\cite{Xia2023AutomatedPR} generate candidate fixes guided by failing tests and repair-specific constraints. These systems primarily study how to implement or validate repairs; in contrast, \swemanager{} studies the earlier managerial step of selecting among competing proposals and synthesizing a final proposal before code modification.

\section{CONCLUSION}
This paper studies proposal selection as a distinct managerial capability in LLM-based software engineering: before code is changed, a system must decide which of several plausible implementation proposals should guide the repair. Our manual study shows that maintainer choices are not arbitrary. They are guided by recurring high-level rationales, made through context-conditioned comparison among proposals, and shaped by discussion that can refine the final implementation direction. These findings motivate formulating proposal selection as a learnable comparison-and-synthesis task rather than an isolated label-prediction problem.

Building on this formulation, we introduce \swemanager{} and instantiate it as \modelName{}, an 8B model trained to select the best proposal, justify the choice, and synthesize a golden proposal. On SWE-Lancer Manager, \modelName{} achieves 53.21\% selection accuracy and a 57.75\% earned rate, surpassing GPT-5 and supporting the claim that proposal-selection policies can be learned by a small model. To evaluate whether the synthesized golden proposal is useful beyond selection, we introduce \frameworkName{}, where proposal agents draft candidates, \swemanager{} synthesizes a golden proposal, and an implementation agent produces the patch. On SWE-Lancer IC, \frameworkName{} with \swemanager{} reaches a 55.6\% pass rate, matching \frameworkName{} with GPT-5 and improving over the strongest direct mini-SWE-agent baseline by 7.1\%. Overall, these results indicate that learned proposal selection and synthesis can improve LLM-based issue resolution, while larger proposal pools and longer contexts remain important directions for future work.

\section{Data Availability}
We provide the code, data-processing scripts, training and evaluation scripts needed to reproduce the reported results in the public artifact repository: \url{https://github.com/shuaijiumei/SWE-Manager}.

%% file: input/selector_comparison_among_proposal_num.tex
\begin{figure}[!htp]
  \centering
  \makebox[\linewidth][c]{\includegraphics[width=0.85\linewidth]{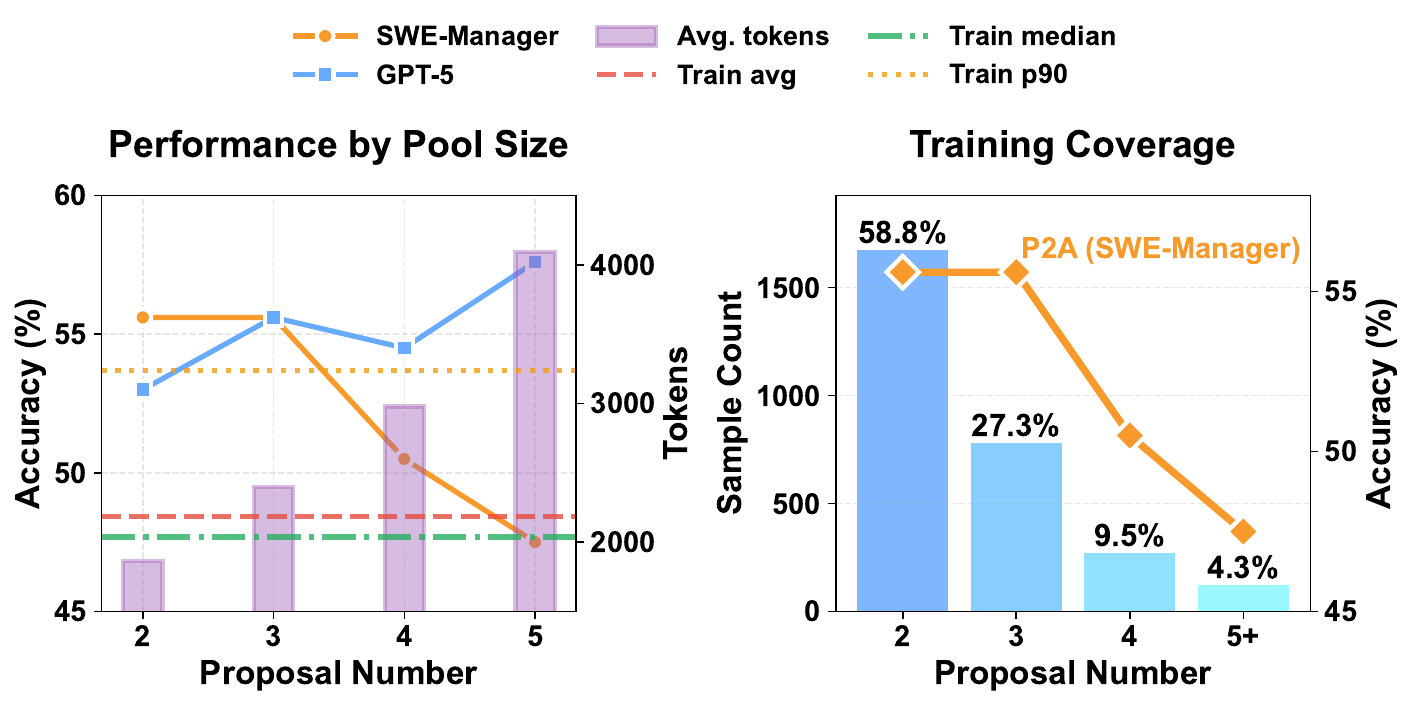}}
  \Description{The left panel plots manager performance and average input length as candidate-pool size increases. The right panel relates the distribution of training candidate counts to downstream performance with SWE-Manager.}

\caption{Effect of proposal-pool size on selector robustness. The left panel reports P2A performance as the number of candidate proposals increases, together with average input length. The right panel compares training-set proposal-count coverage with P2A performance using \modelName{}.}

  \label{fig:rq2_ablation}
\end{figure}

%% file: input/Table_related_selection.tex
\begin{table*}[t]
\centering
\caption{How LLM-based software engineering systems handle multiple candidates for the same task.}
\label{tab:related_selection}
\small
\setlength{\tabcolsep}{4pt}
\renewcommand{\arraystretch}{1.15}
\renewcommand{\tabularxcolumn}[1]{m{#1}}
\begin{tabularx}{\textwidth}{@{}>{\raggedright\arraybackslash}m{0.17\textwidth} >{\raggedright\arraybackslash}X >{\raggedright\arraybackslash}m{0.15\textwidth} >{\raggedright\arraybackslash}X >{\raggedright\arraybackslash}m{0.15\textwidth}@{}}
\toprule
\textbf{System} & \textbf{Candidates} & \textbf{Selection time} & \textbf{Selection signal} & \textbf{Non-selected candidates} \\
\midrule
MetaGPT, ChatDev, MAGIS & A single plan or design & --- & No selection & --- \\
\midrule
Self-planning & A single natural-language plan & --- & No selection & --- \\
MapCoder & Natural-language plans from the same model & Before and after implementation & Self-reported confidence orders the plans; sample tests accept one & Kept only as fallbacks \\
PlanSearch & Natural-language plans, each implemented as a program & After implementation & pass@$k$, optionally after public-test filtering & Not merged \\
\midrule
Agentless & Sampled patches & After implementation & Regression and reproduction tests, then majority voting & Discarded \\
SWE-Search & Final patches from tree search & After implementation & Multi-agent debate & Discarded \\
\midrule
\textbf{\swemanager{} (ours)} & \textbf{Competing natural-language proposals} & \textbf{Before implementation} & \textbf{Learned from maintainer decisions} & \textbf{Synthesized into a golden proposal} \\
\bottomrule
\end{tabularx}
\end{table*}